\documentstyle{article}
\def\eqnum#1{\eqno (#1)}
\newcommand{\bmat}{\left(\begin{array}}
\newcommand{\emat}{\end{array}\right)}
\newcommand{\cn}{\mbox{cn}}
\newcommand{\dn}{\mbox{dn}}
\newcommand{\sn}{\mbox{sn}}
\newcommand{\ba}{\begin{array}}
\newcommand{\ea}{\end{array}}
\title{
TETRAHEDRON EQUATION
AND THE ALGEBRAIC GEOMETRY}
\author
{I.G.Korepanov
}
\date{}
\begin{document}
\maketitle

{\bf Abstract.} The tetrahedron equation arises as a generalization  of
the famous Yang---Baxter equation to the 2+1-dimensional quantum field
theory and the 3-dimensional statistical mechanics. Very little is
still known about its solutions. Here a systematic method is
described that does produce non-trivial solutions to the tetrahedron
equation with spin-like variables on the links. The essence of the
method is the use of the so-called tetrahedral Zamolodchikov
algebras.

\setcounter{section}{-1}
\section{Introduction}

Considerable successes made in recent years in the 2-dimensional
statistical mechanics and the 1+1-dimensional quantum field
theory are closely connected with the studying of the Yang---Baxter
equation, also called triangle equation, and finding new
solutions to this equation. The tetrahedron equation is an
attempt to approach the 3-dimensional problems in the same way.
It is known now that there do exist nontrivial solutions to the
tetrahedron equation [1,2,5,8,9]. Despite some setbacks in the
existing solutions, such as negativeness of some of their matrix
elements, which does not allow to use them directly in statistical
mechanics models, the tetrahedron equation is no doubt worth
further studying. The guarantee is the mathematical beauty
already revealed in this area.

Properly speaking, there are different modifications of the
tetrahedron equation. Here I study (unlike the papers [1,2,10,11,12])
the equation ``with variables on the links''. This means the
equalness of two operator products in the tensor product of 6
complex linear spaces:
$$S_{123}S'_{145}S''_{246}S'''_{356}= S'''_{356}S''_{246}S'_{145}S_{123}.$$
Each operator acts nontrivially only in the tensor product of 3
spaces numbered by its subscripts. This is illustrated by Fig. 1.

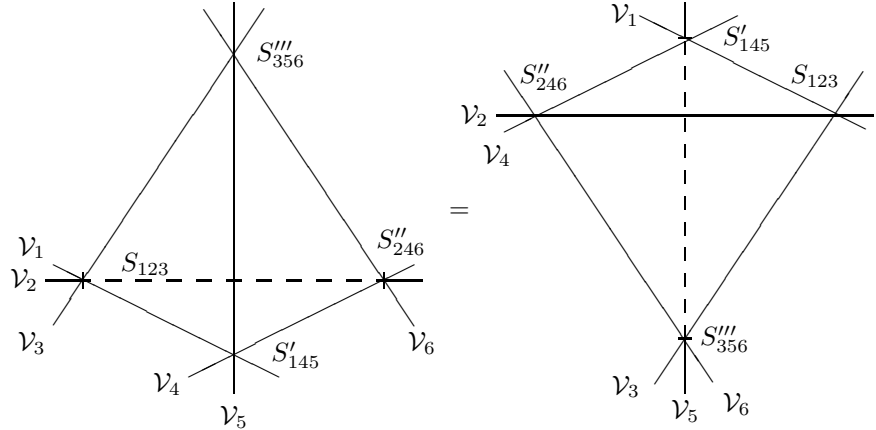
\begin{figure}
\begin{center}
\unitlength=1.00mm
\special{em:linewidth 0.4pt}
\linethickness{0.4pt}
\begin{picture}(115.00,55.92)
\put(115.00,40.86){\line(-1,0){50.00}}
\put(90.00,11.19){\dashbox{1.72}(0.00,40.00)[cc]{}}
\put(90.00,51.19){\line(0,1){4.73}}
\put(114.00,39.14){\line(-2,1){30.00}}
\put(97.00,54.20){\line(-2,-1){31.00}}
\put(86.00,5.17){\line(2,3){27.67}}
\put(66.00,46.89){\line(2,-3){27.67}}
\put(90.00,11.19){\line(0,-1){7.31}}
\put(30.00,55.92){\line(0,-1){52.04}}
\put(5.00,18.93){\line(1,0){5.00}}
\put(50.00,18.93){\line(1,0){5.00}}
\put(10.00,18.93){\dashbox{2.00}(40.00,0.00)[cc]{}}
\put(54.00,12.91){\line(-2,3){28.00}}
\put(34.00,55.06){\line(-2,-3){28.00}}
\put(54.00,21.08){\line(-2,-1){30.00}}
\put(36.00,6.03){\line(-2,1){30.00}}
\put(5.00,22.37){\makebox(0,0)[rb]{${\cal V}_1$}}
\put(4.00,18.93){\makebox(0,0)[rc]{${\cal V}_2$}}
\put(5.00,12.05){\makebox(0,0)[rt]{${\cal V}_3$}}
\put(23.00,5.17){\makebox(0,0)[rc]{${\cal V}_4$}}
\put(30.00,2.16){\makebox(0,0)[ct]{${\cal V}_5$}}
\put(55.00,12.05){\makebox(0,0)[ct]{${\cal V}_6$}}
\put(33.00,49.04){\makebox(0,0)[lc]{$S'''_{356}$}}
\put(15.00,20.22){\makebox(0,0)[lb]{$S_{123}$}}
\put(35.00,9.04){\makebox(0,0)[lc]{$S'_{145}$}}
\put(49.00,23.23){\makebox(0,0)[lb]{$S''_{246}$}}
\put(60.00,27.96){\makebox(0,0)[cc]{$=$}}
\put(65.00,36.99){\makebox(0,0)[ct]{${\cal V}_4$}}
\put(62.00,40.86){\makebox(0,0)[cc]{${\cal V}_2$}}
\put(82.00,54.20){\makebox(0,0)[cc]{${\cal V}_1$}}
\put(84.00,3.88){\makebox(0,0)[rb]{${\cal V}_3$}}
\put(90.00,3.01){\makebox(0,0)[ct]{${\cal V}_5$}}
\put(95.00,3.88){\makebox(0,0)[lt]{${\cal V}_6$}}
\put(68.00,45.17){\makebox(0,0)[lb]{$S''_{246}$}}
\put(95.00,51.19){\makebox(0,0)[lc]{$S'_{145}$}}
\put(104.00,45.17){\makebox(0,0)[lb]{$S_{123}$}}
\put(92.00,11.19){\makebox(0,0)[lc]{$S'''_{356}$}}
\end{picture}
\end{center}
\caption{The diagrammatic representation of the tetrahedron equation
with variables on the links}
\end{figure}

In describing the solutions to the tetrahedron  equation, I
follow the papers [8,9]. The way I came to these solutions was
through the ``tetrahedral Zamolodchikov algebras'' (TZA's) [3]. These
are some structures lying between the triangle equation and the
tetrahedron equation. As has been shown in [3], the TZA's may be
used, at least, for constructing a statistical mechanical model
on the two neighboring layers of the 3-dimensional cubic lattice,
the corresponding Boltzmann weights being positive if the
parameters satisfy some inequalities.

I begin with a study of the well-known Felderhof ${\cal L}$-operators
[6,4] and their products by means of  algebraic geometry (section 1).
This provides a large amount of TZA's (section 2). The solutions to
the tetrahedron equation are presented in section 3. I conclude
with a discussion in section 4.

\section{Vacuum Curves and Vacuum Vectors of the Felderhof
${\cal L}$-Operators' Products}

\subsection{Vacuum curve and vacuum vectors of a Felderhof
${\cal L}$-operator}

In this subsection and partly in the in the next I recall some
facts from [4].
Let ${\cal V}_{0} , {\cal V}_{1} , \ldots
$ be 2-dimensional complex linear spaces with
fixed bases. Felderhof ${\cal L}$-operator is a linear operator acting
in the tensor product of two such spaces, e.g. ${\cal V}_{0} \otimes
{\cal V}_{1}$, of
the following form:
$$L =\bmat{cccc}a_{+}&&&{d}\\
&{b_{-}}&c&\\
&c&{b_{+}}&\\
{d}&&&{a_{-}}
\emat,$$
where

$$
a_+a_-+b_+b_-=c^2+d^2;
$$
each $2\times 2$ block is an operator in ${\cal V}_{1}$. Let us denote the
Felderhof ${\cal L}$-operators by the letters $L,M,\ldots
$ . To emphasize the
spaces in which the operator acts, I will write also $L = L_{01}$ etc.
It will be useful to consider the vectors from the spaces ${\cal V}_{k}$
with the second coordinate equal to unity. Let us denote them as
follows:
$$U =\bmat{c}u\\1\emat,\quad V =\bmat{c}v\\1\emat,
\quad X =\bmat{c}x\\1\emat.$$
Let $U, V \in  {\cal V}_{0},\; X,Y \in  {\cal V}_{1}$.   Under what
conditions does the equality
$$L(U \otimes  X) = g(V \otimes  Y)\eqno(1)$$
hold ($ g$  being a numerical factor)?
The condition turns out to be the following connection
between $u$ and $v$:
$$u^{2}v^{2}+ \alpha u^{2} + \delta v^{2} + 1 = 0, \eqno(2)$$
where
$$\alpha  = {a_{+}b_{-}\over cd} ,\quad \delta  = {a_{-}b_{+}\over cd}.$$
Eq. (2) determines an elliptic curve $\Gamma _{L}$ , and to each point
$(u,v) \in  \Gamma _{L}$ corresponds a 1-dimensional space of the vectors
proportional to $X$ . So the vectors proportional to $X$---let us call
them the vacuum vectors of the operator $L$---make up a 1-dimentional
vector bundle over $\Gamma _{L}$. This bundle, as usually, may be
determined by
a divisor---e.g.\ by the pole divisor of the vector $X$ first
coordinate $x$ . Denote this divisor $D_{L}$. One can find from Eq. (1)
that it consists of 2 points $(u,v) = (u_{L},v_{L})$ and $(-u_{L},-v_{L})$,
where
$$u^{2}_{L}= {cb_{+}\over da_{+}} ,\quad v_{L} = {a_{+}\over b_{+}} u_{L}.$$
When necessary, I will add an index: $X = X_{L},\; x = x_{L}$ etc.
\par
\medskip

\subsection{The vacuum curve and the vacuum vector bundle of the
product of two Felderhof ${\cal L}$-operators}

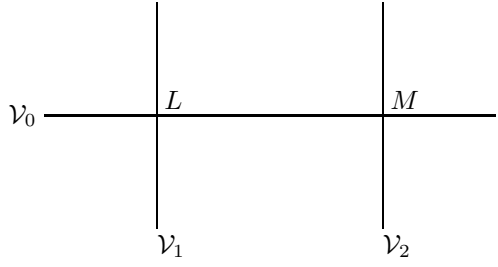
\begin{figure}
\begin{center}
\unitlength=1.00mm
\special{em:linewidth 0.4pt}
\linethickness{0.4pt}
\begin{picture}(63.00,33.12)
\put(3.00,18.06){\line(1,0){60.00}}
\put(18.00,33.12){\line(0,-1){30.11}}
\put(48.00,3.01){\line(0,1){30.11}}
\put(2.00,18.06){\makebox(0,0)[rc]{${\cal V}_0$}}
\put(18.00,2.15){\makebox(0,0)[lt]{${\cal V}_1$}}
\put(48.00,2.15){\makebox(0,0)[lt]{${\cal V}_2$}}
\put(19.00,18.92){\makebox(0,0)[lb]{$L$}}
\put(49.00,18.92){\makebox(0,0)[lb]{$M$}}
\end{picture}
\end{center}
\caption{The diagrammatic representation of the product of two
$\cal L$-operators}
\end{figure}

Consider now two Felderhof ${\cal L}$-operators $L = L_{01}$ and $M = M_{02}$.
Consider their product $LM$ acting in ${\cal V}_{0}\otimes
{\cal V}_{1}\otimes  {\cal V}_{2}$ (Fig. 2). Let us
investigate the vacuum vectors ${\cal X}_{LM}\in  {\cal V}_{1}\otimes
{\cal V}_{2}$   satisfying the relation
$LM(U \otimes  {\cal X}_{LM}) = g ( V \otimes  {\cal Y}_{LM})$.
Dependence between $u$ and $v$ is found according to [4]. Let us
write down the equations of the vacuum curves $\Gamma _{L}$ and
$\Gamma _{M}$ , and
denote their points in the following way: $\Gamma _{L} \ni  (w,v),\;
\Gamma _{M} \ni  (u,w)$:
$$
\Gamma _{L}:\qquad w^{2}v^{2}+ \alpha _{L}w^{2} + \delta _{L}v^{2} + 1 = 0,
\eqno(3)
$$ $$
\Gamma _{M}:\qquad u^{2}w^{2}+ \alpha _{M}u^{2} + \delta _{M}w^{2} + 1 = 0.
\eqno(4)
$$
The vacuum curve equation of the operator $LM$ is found by eliminating
$w^{2}$ from Eqs. (3,4):
\par
$$
\Gamma _{LM}:\; (\delta _{L}-\alpha _{M})u^{2}v^{2} + (1-\alpha _{L}
\alpha _{M})u^{2} + (\delta _{L}\delta _{M}-1)v^{2} + (\delta _{M}-
\alpha _{L})=0.
\eqnum{5}$$
\noindent This time a 2-dimensional space of vectors ${\cal X}_{LM}$
corresponds to each
point $(u,v)\in \Gamma _{LM}$, as one can see in Fig.~3.
\begin{figure}
\begin{center}
\unitlength=1.00mm
\special{em:linewidth 0.4pt}
\linethickness{0.4pt}
\begin{picture}(106.00,48.17)
\put(65.00,0.00){\line(-2,1){60.00}}
\put(5.00,30.00){\line(6,1){60.00}}
\put(65.00,40.00){\line(2,-1){40.00}}
\put(105.00,20.00){\line(-2,-1){40.00}}
\put(4.00,30.00){\makebox(0,0)[rc]{$u$}}
\put(65.00,40.86){\makebox(0,0)[cb]{$w$}}
\put(65.00,2.15){\makebox(0,0)[cb]{$-w$}}
\put(106.00,19.78){\makebox(0,0)[lc]{$v$}}
\put(46.00,37.85){\makebox(0,0)[rb]{$X_M(u,w)$}}
\put(29.00,18.92){\makebox(0,0)[lb]{$X_M(u,-w)$}}
\put(77.00,34.84){\makebox(0,0)[lb]{$X_L(w,v)$}}
\put(89.00,12.04){\makebox(0,0)[rb]{$X_L(-w,v)$}}
\put(35.00,48.17){\makebox(0,0)[cc]{\large$\Gamma_M$}}
\put(85.00,48.17){\makebox(0,0)[cc]{\large$\Gamma_L$}}
\end{picture}
\end{center}
\caption{Vacuum curve and vacuum vectors of the product of
two ${\cal L}$-operators   }
\end{figure}
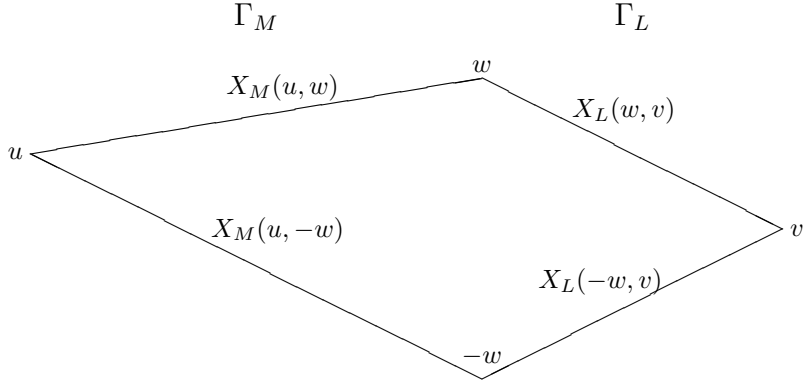

The 2-dimensional space
contains the vectors
$$
X_{L}(w,v) \otimes  X_{M}(u,w)
\eqnum{6}$$
and
$$
X_{L}(-w,v) \otimes  X_{M}(u,-w).
\eqnum{7}$$

{\bf Theorem 1.} {\it The vacuum vector bundle of the operator $LM$
decomposes into a direct sum of two sub-bundles ${\cal E}_{1}$ and
${\cal E}_{2}$ of degree 2.
The bundles ${\cal E}_{1}$ and ${\cal E}_{2}$ are, generally,
non-isomorphic, yet each of them
becomes isomorphic to the other after the transform
$(u,v)\rightarrow (u,-v)$ of
the curve $\Gamma _{LM}$.
}\smallskip\\

{\it Proof.} Rewrite the vectors (6,7) together as a matrix
$$
H = \bmat{cc}x^{+}_{L} x^{+}_{M} & x^{-}_{L} x^{-}_{M} \smallskip
\\
x^{+}_{L}& x^{-}_{L}\smallskip\\x^{+}_{M}& x^{-}_{M}\smallskip\\1&1
\emat
$$
 where I use brief notations
$$
X_{L}(\pm w,v)=\bmat{c}x^{\pm}_L\\1 \emat,
\qquad X_{M}(u,\pm w)=\bmat{c}x^{\pm}_{M}\\1 \emat.
$$
One can get two other vectors forming a basis in the fiber by multiplying
$H$ from the right by any non-degenerate $2\times2$ matrix $G$. Let us
choose
$$G = \bmat{cc}x^{+}_{L}x^{+}_{M}& x^{-}_{L}x^{-}_{M}  \\
1&1 \emat^{-1}.
$$

We find then
$$
HG = \bmat{ccc}1&&0\\ &{\bf K}&\\ 0&&1 \emat,
$$
where
$$
{\bf K} = (x^{+}_{L}x^{+}_{M} - x^{-}_{L}x^{-}_{M})^{-1}\cdot
\bmat{cc}x^{+}_{L} - x^{-}_{L}& x^{+}_{L}x^{-}_{L}
(x^{+}_{M}-x^{-}_{M}) \smallskip \\
x^{+}_{M} - x^{-}_{M} & x^+_M x^-_M (x^{+}_{L}-x^{-}_{L})
\emat
$$

Thus, the matrix $HG$ remains invariant under the change $w \leftrightarrow
-w$ and
depends only on $u$ and $v,\; (u,v)\in  \Gamma _{LM}$. Now investigate its
matrix
elements' poles. Thorough examination of all ``suspicious'' points shows
that elements of the matrix ${\bf K}$ have poles in the points where
$x^{+}_{L}x^{+}_{M} - x^{-}_{L}x^{-}_{M} = 0$ except the points with
$w=0$ or $\infty$.

A routine consideration shows the degree of the matrix ${\bf K}$ pole
divisor to equal 4, the poles being located in some points
$$
(u_{1},v_{1}),\; (-u_{1},-v_{1}),\; (u_{2},v_{2}),\; (-u_{2},-v_{2}).
\eqnum{8}$$
Let the ratios of the residues of the matrix ${\bf K}$ first column
elements to
those of the second column be $\beta :1$ in the first two points (8) and
$\gamma :1$
in the others (that is what is called a matrix divisor, see [4,7]).
Multiply $HG$ from the right by a matrix
$$
G_{1} = \bmat{cc}1&1 \\ -\gamma & -\beta \emat.
$$
Then the first column of $HGG_{1}$ has only two poles $(u_{1},v_{1})$ and
$(-u_{1},-v_{1})$,
i.e. its pole divisor $D_{1}= (u_{1},v_{1})+(-u_{1},-v_{1})$. Similarly,
the second
column has a pole divisor $D_{2}=(u_{2},v_{2})+(-u_{2},-v_{2})$. Evidently,
 ${\cal E}={\cal E}_{1}\oplus  {\cal E}_{2}$,
the bundles ${\cal E}_{1}$ and ${\cal E}_{2}$ corresponding to the
divisors $D_{1}$ and $D_{2}$.

To prove that ${\cal E}_{1}$ (or ${\cal E}_{2})$ transforms into a bundle
isomorphic
to ${\cal E}_{2}$ (or ${\cal E}_{1})$ by the isomorphism $(u,v)\rightarrow
(u,-v)$ of the curve $\Gamma _{LM}$, let us
multiply the matrix $H$ from the right by a matrix
$$G' = \bmat {cc} x^{+}_{M}& x^{-}_{M} \\ 1&1 \emat $$
Again the elements of the matrix $HG'$ have 4 poles, but now those poles
coincide with the poles of $x_{L}$. The 4 corresponding points are of the
form $(\pm u_{0},\pm v_{0})$ with arbitrary signs chosen independently. The
divisor $D$ of those poles can now be seen to be equivalent to
$2D_{u}, \quad D_{u}$
being a pole divisor of the function $u$ on $\Gamma _{LM}$.

The divisor $D$ corresponds to the determininant (see e.g. [7])
of the bundle ${\cal E}$, thus $D \sim  D_{1}+ D_{2}$, i.e.
$$
D_{1} + D_{2}\sim 2D_{u} .
\eqnum{9}$$
One can easily see that for each divisor $D_{1}$ of degree 2
$$
D_{1} + \tilde{D}_{1} \sim  2D_{u} ,\eqno(10)
$$
where $\tilde{D}_{1}$ is obtained from $D_{1}$ by changing
$v\rightarrow -v$. From Eqs (9,10) we
find $D_{2}\sim  \tilde{D}_{1}$, and with this Theorem 1 is completely
proved.

\subsection{Vacuum vector bundle of the product of 3 Feldergof
${\cal L}$-operators
}

Consider now a product $LMN= L_{01}M_{02}N_{03}$ of 3 Felderhof
${\cal L}$-operators.
\smallskip

{\bf Theorem 2.} {\it The vacuum vector bundle ${\cal F}$ of the operator
$LMN$ is a
direct sum ${\cal F} = {\cal F}_{1}\oplus  {\cal F}_{2} \oplus
{\cal F}_{3}\oplus  {\cal F}_{4}$, where ${\cal F}_{1} \sim  {\cal F}_{2},
 {\cal F}_{3} \sim  {\cal F}_{4}$, but ${\cal F}_{1}$ is,
generally,  not isomorphic to ${\cal F}_{3}$.
}\smallskip\\
{\it Proof.} According to Theorem 1, the vacuum vectors of the
operator $LM$ form the 2-dimensional bundle ${\cal E} = {\cal E}_{1}\oplus
{\cal E}_{2}$, and each of
${\cal E}_{1}$ and ${\cal E}_{2}$ is of the same kind as if it were a
vacuum vector bundle of
some Felderhof ${\cal L}$-operator. Having multiplied $LM$ by $N$, one
obtains a
2-dimensional bundle, say ${\cal F}_{1}\oplus  {\cal F}_{3}$, out of
 the vectors of ${\cal E}_{1}$, and
another 2-dimensional bundle, say ${\cal F}_{2}\oplus  {\cal F}_{4}$, out
of the vectors of ${\cal E}_{2}$.
Under the transform $(u,v)\rightarrow (u,-v)$ of the curve
$\Gamma _{LMN}$, the bundles
transform, up to the isomorphisms, as follows:
$$
{\cal F}_{1} \leftrightarrow  {\cal F}_{3},\quad {\cal F}_{2}
\leftrightarrow  {\cal F}_{4}
\eqnum{11}$$
On the other hand, under the same transform of the curve $\Gamma _{LM}$,
${\cal E}_{1}\leftrightarrow  {\cal E}_{2}$, and consequently
$({\cal F}_{1}\oplus  {\cal F}_{3})\leftrightarrow  ({\cal F}_{2}\oplus
  {\cal F}_{4})$. One may assume
that
$$
{\cal F}_{1}\leftrightarrow  {\cal F}_{4},\quad {\cal F}_{2}
\leftrightarrow  {\cal F}_{3}.
\eqnum{12}$$
Combine the formulae (11, 12), and the proof is completed.
\smallskip

{\bf Corollary.} {\it The linear space of endomorphisms of the bundle
${\cal F}$
is 8-dimen\-sional.
}

{\it Proof.} One can see from the Theorem 2 that each fiber of the
bundle ${\cal F}$  may be generated by linearly independent meromorphic
vectors ${\cal X}_{1}, {\cal X}_{2}, {\cal X}_{3}, {\cal X}_{4}$, each of
the pairs ${\cal X}_{1}, {\cal X}_{2}$ and ${\cal X}_{3}, {\cal X}_{4}$
having a
common pole divisor of degree 2. In the basis $({\cal X}_{1},{\cal X}_{2},
{\cal X}_{3},{\cal X}_{4})$, the
endomorphisms of the bundle ${\cal F}$ are given by arbitrary constant
matrices of the form
$$ \bmat {cc} \ba{cc} a_{11}& a_{12}\\ a_{21} & a_{22}\ea &
\mbox {\raisebox{-1ex}{\huge 0}}\\
\mbox {\raisebox{-1ex}{\huge 0}} & \ba{cc} a_{33} & a_{34} \\
a_{34} & a_{44} \ea \emat
$$
acting from the right.

\section{Tetrahedral Zamolodchikov Algebras}

Now let the Felderhof ${\cal L}$-operators $L_{01}$ and $M_{02}$ satisfy the
Yang---Baxter equation:
$$
R_{12}L_{01}M_{02} = M_{02}L_{01}R_{12}.
\eqnum{13}$$
The matrix $R_{12}$ is then known to be symmetrical: $R^{T}_{12} = R_{12}$.
 However
(see, e.g., [4]), along with $R_{12} = R^{0}_{12}$ there exists a
non-symmetrical matrix $R^{1}_{12}$ such that
$$
(R^{1}_{12})^{T}L_{01}M_{02} = M_{02}L_{01}R^{1}_{12}.
$$
Examples are given in Section 3.

Let the pairs of operators $L_{01},N_{03}$ and $M_{02},N_{03}$  also satisfy
equations of the kind of Eq. (13):
$$
\tilde{R}_{13}L_{01}N_{03} = N_{03}L_{01}\tilde{R}_{13},
$$
$$
\tilde{\tilde{R}}_{23}M_{02}N_{03} = N_{03}M_{02}\tilde{\tilde{R}}_{23}.
$$
Here the tildes indicate that the $R$-operators not only act in
different spaces but differ in their matrix elements. Somewhat
freely, I will allow myself to omit those tildes. Then I will
${\rm re}$-denote $R_{13} = R^{0}_{13},\; R_{23} = R^{0}_{23}$, and
introduce $R^{1}_{13}$ and $R^{1}_{23}$ as before.
Let us consider the operators ${\cal R} = {\cal R}_{123},\;
\hat{{\cal R}} = \hat{{\cal R}}_{123}$ which permute the
${\cal L}$-operators as follows:
$$
\hat{{\cal R}}_{123}L_{01}M_{02}N_{03} = N_{03}M_{02}L_{01}{\cal R}_{123}.
\eqnum{14}$$
One gets 8 such ${\cal R}$-operators (generally, linearly independent---see
 Section 4) in the form
$$
{\cal R} = R^{a}_{12}R^{b}_{13}R^{c}_{23}
$$
with $a,b,c=0$ or 1. The corresponding $\hat{{\cal R}}$-operators are
$$
\hat{{\cal R}} = (R^{c}_{23}R^{b}_{13}R^{a}_{12})^{T}
$$
(note that in [8,9] it was stated that $\hat{{\cal R}}={\cal R}^{T}$,
which is a mistake). On
the other hand, there are also 8 ${\cal R}$-operators of the form
$$
{\cal R} = R^{f}_{23}R^{e}_{13}R^{d}_{12},
$$
with corresponding
$$
\hat{{\cal R}} = (R^{d}_{12}R^{e}_{13}R^{f}_{23})^{T}.
$$
It follows from Eq.~(14) that ${\cal R}_{123}$ converts the vacuum vectors
of $L_{01}M_{02}N_{03}$ into those of $N_{03}M_{02}L_{01}$. The
corresponding vacuum
vector bundles are isomorphic. Thus, it follows from the Corollary
of Theorem 2 that the linear space of the operators ${\cal R}_{123}$ is
8-dimensional. This leads to the linear dependences of a tetrahedral
Zamolodchikov algebra (Ref. 3):
$$
R^{a}_{12}R^{b}_{13}R^{c}_{23} =\sum^{1}_{d,e,f=0} S^{abc}_{def}
R^{f}_{23}R^{e}_{13}R^{d}_{12} .
\eqnum{15}$$
\section{New Solutions to the Tetrahedron Equation}

Consider a Felderhof ${\cal L}$-operator of the form
$$
L(\lambda ) = \bmat{cccc} \cn \lambda &&&k\, \sn\lambda    \\
& \sn\lambda  \dn \lambda & \dn\lambda  & \\
&\dn\lambda  & \sn\lambda \dn  \lambda & \\
k\,\sn \lambda  \cn \lambda &&& \cn \lambda      \emat
$$
$k$ being a modulus of elliptic functions. Let us fix 4 complex
numbers $\lambda  = \lambda _{1},\lambda _{2},\lambda _{3},\lambda _{4}$,
and consider 4 operators
$$
L_{01}(\lambda _{1}),\; L_{02}(\lambda _{2}),\; L_{03}(\lambda _{3}),\;
L_{04}(\lambda _{4}).
$$
Sometimes I will omit the arguments $\lambda _{j}$. Let us introduce the
operators \\
$R^{e}_{ij}(\lambda _{i}, \lambda _{j}),$ $1\leq i<j\leq 4,$
$e=0,1,$ by the following formulae:
$$
R^{0}_{ij}(\lambda _{i}, \lambda _{j}) = f_{0}(\lambda _{i},
\lambda _{j})\bmat{cccc} a&&&d\\&b&c&\\&c&b&\\d&&&a \emat,
$$
$$
R^{1}_{ij}(\lambda _{i}, \lambda _{j}) = f_{1}(\lambda _{i},
\lambda _{j}) \bmat{cccc} -a'&&&-d'\\ &-b'& c'&\\
&-c'&b'&\\ -d'&&&a'\emat.
$$
Herein
$$
a=\cn(\lambda _{i}- \lambda _{j}),\quad b=\sn(\lambda _{i}- \lambda _{j})\dn
(\lambda _{i}- \lambda _{j}),
$$
$$
c=\dn(\lambda _{i}- \lambda _{j}),\quad d=k\sn
(\lambda _{i}- \lambda _{j})
\cn(\lambda _{i}- \lambda _{j}),
$$
$$
a'=\cn(\lambda _{i}+ \lambda _{j}), \quad b'=\sn(\lambda _{i}+
\lambda _{j})\dn
(\lambda _{i}+ \lambda _{j})
$$
$$
c'=\dn(\lambda _{i}+ \lambda _{j}),\quad
d'=k\sn(\lambda _{i}+ \lambda _{j})\cn
(\lambda _{i}+ \lambda _{j}),
$$
$f_{0}(\lambda _{i},\lambda _{j})$ and $f_{1}(\lambda _{i},\lambda _{j})$
are arbitrary numerical factors. One can
verify that the equations of  the Section 3 hold:
$$
R^{0}_{12}L_{01}L_{02}=L_{02}L_{01}R^{0}_{12}, \quad
(R^{1}_{12})^{T} L_{01}L_{02}
= L_{02}L_{01}R^{1}_{12}
$$
etc.

Now let us introduce the 2-dimensional linear spaces ${\cal V}_{12}$,
${\cal V}_{13}$,
${\cal V}_{14}$, ${\cal V}_{23}$, ${\cal V}_{24}$, ${\cal V}_{34}$
and consider the
 matrix $S$ from Eq.(15) as a
linear operator in ${\cal V}_{12}\otimes  {\cal V}_{13}\otimes
{\cal V}_{23}$, so that the indices $a$ and $d$
correspond to the space ${\cal V}_{12}$ and so on. It can be verified
that for
generic $k$ and $\lambda _{j}$ the matrix $S$ is determined from (15)
uniquely,
unlike the special case of the paper $[3]$. I will write
$$
S = S_{12,13,23} = S_{12,13,23}(\lambda _{1},\lambda _{2},\lambda _{3};k).
$$
Consider then 3 more $S$-matrices
$S_{12,14,24},\; S_{13,14,34},\; S_{23,24,34}$,
whose definition is obvious (so that, e.g., $S_{12,14,24}$  acts in
${\cal V}_{12}\otimes  {\cal V}_{14}\otimes  {\cal V}_{24}$  and depends on
$\lambda _{1}, \lambda _{2}, \lambda _{4}; k )$. Does the tetrahedron
equation
$$
S_{12,13,23}S_{12,14,24} S_{13,14,34}S_{23,24,34} =
S_{23,24,34}S_{13,14,34}S_{12,14,24}S_{12,13,23}\eqno(16)
$$
hold?

The answer is positive, but the calculations for generic $k$ are
rather difficult and will be presented elsewhere. In this paper, I
will restrict myself to the case $k\rightarrow 0,$ in which I have actually
obtained the matrix $S$ directly from Eq.~(15) (one cannot just take
$k=0$ because of the linear dependence of the corresponding
$R$-operators' products---see [3]). For the elements of the $S$-matrices
to have the simplest form, let us choose
$$
f_{0}(\lambda _{i} , \lambda _{j}) = \sin ^{-1}(\lambda _{i} - \lambda _{j}),
$$
$$
f_{1}(\lambda _{i} , \lambda _{j}) = \cos ^{-1}(\lambda _{i} + \lambda _{j})
$$
and introduce new variables $\varphi _{1}, \varphi _{2}, \varphi _{3},
\varphi _{4}$ by equalities
$$
\tanh  \varphi _{j} = \tan ^{2}\lambda _{j}.
$$
The matrix elements of $S=S_{12,13,23}$ are (for other matrices, make the
obvious change of indices):
$$
S^{000}_{000} = S^{011}_{011} = S^{101}_{101} = S^{110}_{110} = 1,
$$
$$
S^{001}_{010} =\hbox{ cotanh}(\varphi _{1} - \varphi _{3})\tanh
(\varphi _{2} - \varphi _{3}),
$$
$$
S^{001}_{100} = \tanh (\varphi _{2} - \varphi _{3}),
$$
$$
S^{001}_{111} =\hbox{ cotanh}(\varphi _{1} - \varphi _{3}),
$$
$$
S^{010}_{001} = S^{010}_{100} = S^{010}_{111} = 1,
$$
$$
S^{100}_{001} = -\tanh (\varphi _{1} - \varphi _{2}),
$$
$$
S^{100}_{010} =\tanh (\varphi _{1} - \varphi _{2})\hbox{cotanh}(\varphi _{1}
 - \varphi _{3}),
$$
$$
S^{100}_{111} = -\hbox{cotanh}(\varphi _{1} - \varphi _{3}),
$$
$$
S^{111}_{001} = \tanh (\varphi _{1} - \varphi _{2}),
$$
$$
S^{111}_{010} = -\tanh (\varphi _{1} - \varphi _{2})\tanh (\varphi _{2} -
 \varphi _{3}) ,
$$
$$
S^{111}_{100} = -\tanh (\varphi _{2} - \varphi _{3}),
$$
all the other matrix elements are zeros.

The way I have proved that these $S$-matrices really satisfy the
tetrahedron equations (16) was as follows. It can be seen that in the
variables
$$
x_{j}=\tanh  \varphi _{j},\quad j=1,\ldots
4,
$$
all the matrix elements of the $S$-matrices become rational functions
with {\it integer} coefficients. After having been multiplied by its
common denominator, Eq.(16) becomes an equality between two matrix
polynomials in $x_{1},\ldots
 x_{4}$ with integer coefficients. Thus, it can be
{\it exactly} verified by a computer, and that is what actually has been
done. More ``scientific'' way of proving Eq. (16) will be presented
elsewhere.

To conclude this section, note that $S$ depends only on the
differences of the arguments $\varphi _{j}$. Less obvious observation is
that $S$
is a reflection: $S^{2}=1,$ with 2 eigenvalues equalling $ -1$.

\section{Discussion}

It is shown in this paper that the tetrahedral Zamolodchikov
algebras do lead to new solutions of the tetrahedron equation with
variables on the links, at least in the particular case of Section~3. It
would be very interesting to calculate the $S$-matrices for
general Felderhof ${\cal L}$-operators and to verify whether they satify the
tetrahedron equation.

A certain disadvantage in the solutions of the tetrahedron
equation known so far is that if their matrix elements are chosen to
be real, some of them become incurably negative. This means that
they cannot be directly used as Boltzmann weights of a statistical
mechanical model on the cubic lattice. However, the experience
obtained from studying the triangle equation suggests that the
tetrahedron equation may as well have the most unexpected
applications in various fields of mathematics.
\par
\medskip
\centerline{REFERENCES
}
\medskip
\noindent 1. Zamolodchikov A.B. Tetrahedron equations and the relativistic
$S$-matrix of straight-strings in 2+1-dimensions // Commun. Math.
Physics. - 1981.V.79.- P. 489-505.
\par
\noindent 2. Baxter R.J. On Zamolodchikov's solution of the tetrahedron
equations // Commun. Math. Physics. - 1983. V.88, No.2. - P.
185-205.
\par
\noindent 3. Korepanov I.G. Tetrahedral Zamolodchikov algebra and the
two-layer flat model in statistical mechanics // Modern Phys. Lett.
B. - 1989. V.3, No 3. - P. 201-206.
\par
\noindent 4. Krichever I.M. Baxter equations and the algebraic geometry //
Funkc. analiz i pril. - 1981. V. 15, No. 2. - P. 22-35 (Russian).
\par
\noindent 5. Bazhanov V.V., Stroganov Yu.G. Free fermions on a
three-dimensional lattice and tetrahedron equations // Nucl. Phys. B.
- 1984. V.~B230[FS10], No.4. - P.435-454.
\par
\noindent 6. Felderhof B.U. Diagonalization of the transfer matrix of the
fermion model //Physica. - 1973. V. 66, No 2. - P. 279-298.
\par
\noindent 7. Tyurin A.N. Classification of the vector bundles over algebraic
curves // Izvestia AN SSSR, ser. matem. - 1965. V. 29. - P. 658-680
(Russian).
\par
\noindent 8. Korepanov I.G. New solutions to the tetrahedron equation /
Chelyabinsk, 1989. - 8p. - Dep. in the VINITI No. 1751-V89 (Russian).
\par
\noindent 9. Korepanov I.G. Applications of the algebro-geometrical
constructions to the triangle and tetrahedron eqations. Ph.D.Thesis
/Leningrad: LOMI, 1989. - 87 p. (Russian).
\par
\noindent 10. Bazhanov V.V., Baxter R.J. New solvable lattice models in three
dimensions. Preprint, 1992. Submitted to J. Stat. Phys.
\par
\noindent 11. Bazhanov V.V., Baxter R.J. Star-triangle relation for a three
\par
\noindent dimensional model. Preprint, 1992. Submitted to J. Stat. Phys.
\par
\noindent 12. Kashaev R.M., Mangazeev V.V., Stroganov Yu.G. Spatial symmetry,
local integrability and tetrahedron equations in the Baxter-Bazhanov
Model. IHEP Preprint 92-63. - Protvino, 1992. - 16 p.
\end{document}